\documentclass[times,twocolumn,final,authoryear]{elsarticle}

\usepackage{jasr}
\usepackage{framed,multirow}

\usepackage{amssymb}
\usepackage{latexsym}
\usepackage{natbib}
\usepackage{adsbib}

\usepackage{url}
\usepackage{xcolor}
\definecolor{newcolor}{rgb}{.8,.349,.1}

\usepackage{lineno}

\usepackage[citebordercolor=white]{hyperref}

\journal{Advances in Space Research}

\begin{document}

\verso{Sanja Danilovic}

\begin{frontmatter}

\title{Modeling of Chromospheric Features and Dynamics in Solar Plage}

\author[ ]{Sanja Danilovic}
\ead{sdani@astro.su.se}

\address[ ]{Institute for Solar Physics, Dept. of Astronomy, Stockholm University, Albanova University Center, 10691 Stockholm, Sweden}

\received{2022}
\finalform{2022}
\accepted{2022}
\availableonline{2022}
\communicated{S. Sarkar}

\begin{abstract}
The chromosphere is a dynamic and complex layer where all the relevant physical processes happen on very small spatio-temporal scales. A few spectral lines that can be used as chromospheric diagnostics give us convoluted information that is hard to interpret without realistic theoretical models. What are the key ingredients that these models need to contain? The magnetic field has a paramount effect on chromospheric structuring. This is obvious from the ubiquitous presence of chromospheric dynamic fibrilar structures visible on the solar disk and at the limb. The numerical experiments presented in this manuscript illustrate the present state of modeling. They showcase to what extent our models reproduce various chromospheric features and their dynamics. The publication describes the effect different ingredients have on chromospheric models and provides a recipe for building one-to-one models. Combining these models with observations will provide insight into the physical processes that take place in the solar atmosphere.
\end{abstract}


\end{frontmatter}


\section{Introduction}
\label{sec1}
The solar atmosphere is complex, non-homogeneous and highly dynamic.~It encompasses plasma widely ranging in density, temperature, ionization and magnetodynamic properties which complicates the modeling. The density drops exponentially with height by 10 magnitudes from near-surface convection layers to the corona. This affects the state of the matter which we aim to describe as realistically as possible in our numerical models \citep{2020LRSP...17....3L}. 

A prime agent causing non-uniformity is magnetism. Looking at the chromosphere, e.g.~in H$\alpha$ line core, the solar surface is covered with opaque long fibrils \citep{2007ASPC..368...27R} that emanate from large magnetic features organized at mesogranular to supergranular scales \citep{2012ApJ...753L..13S}. Their appearance varies depending on which chromospheric line is used for observations and what band-pass range is sampled \citep{2008A&A...480..515C, 2020A&A...637A...1K}. Although the long fibrils tend to outline the direction of the magnetic field in the chromosphere \citep{2022A&A...662A..88V}, their appearance largely depends on the unsigned flux density of the region. So, on the one hand, in active regions and around strong network structures, they are solid, opaque and show large contrast. In quiet Sun regions, on the other hand, they appear scarcer and extend less far from the network where they originate.
\begin{figure}
  \centering
  \includegraphics[width=0.8\linewidth,trim= 0.3cm 13.4cm 0cm 0.2cm,clip=true]{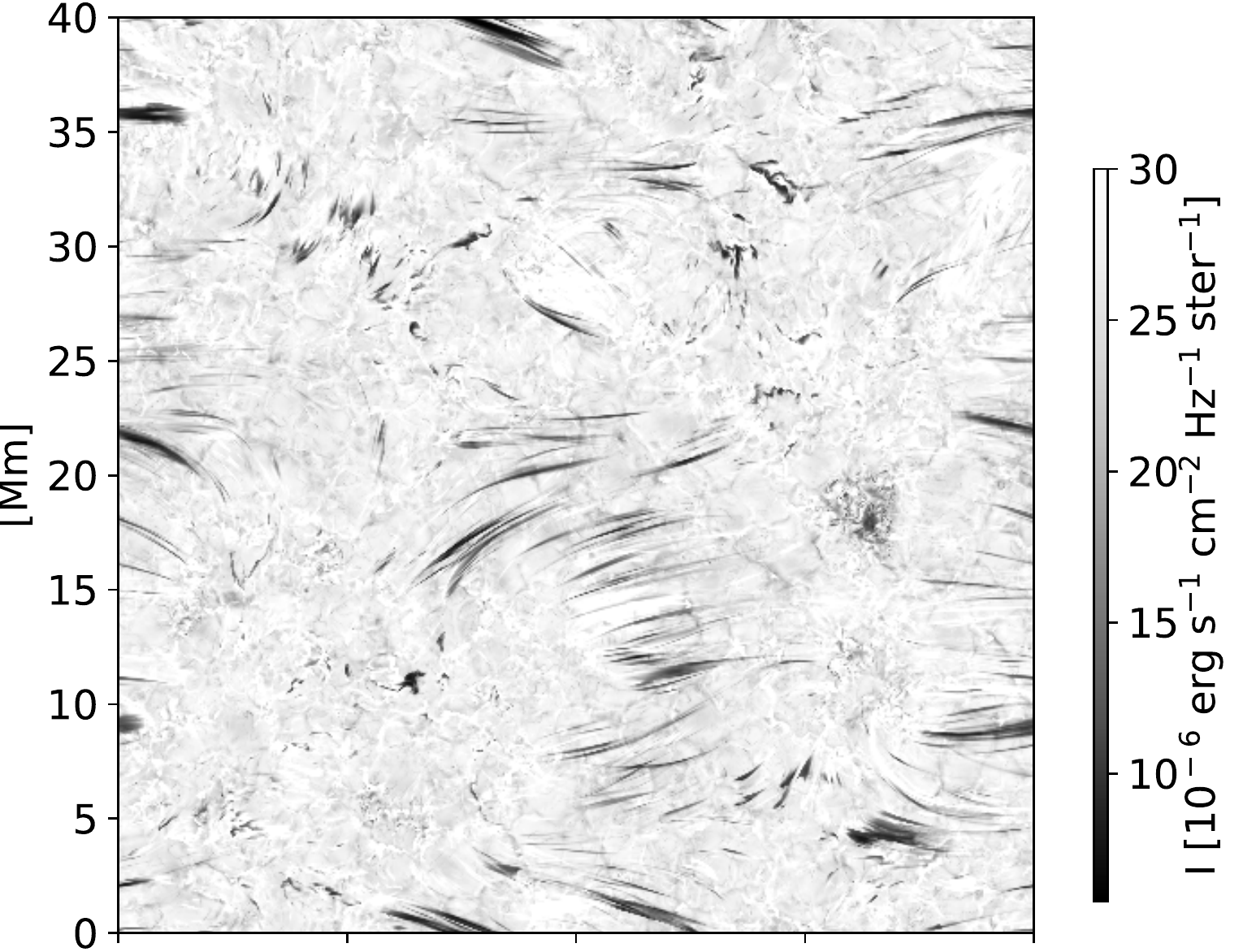}
  \includegraphics[width=0.8\linewidth,trim= 0.3cm 0cm 0cm 0.2cm,clip=true]{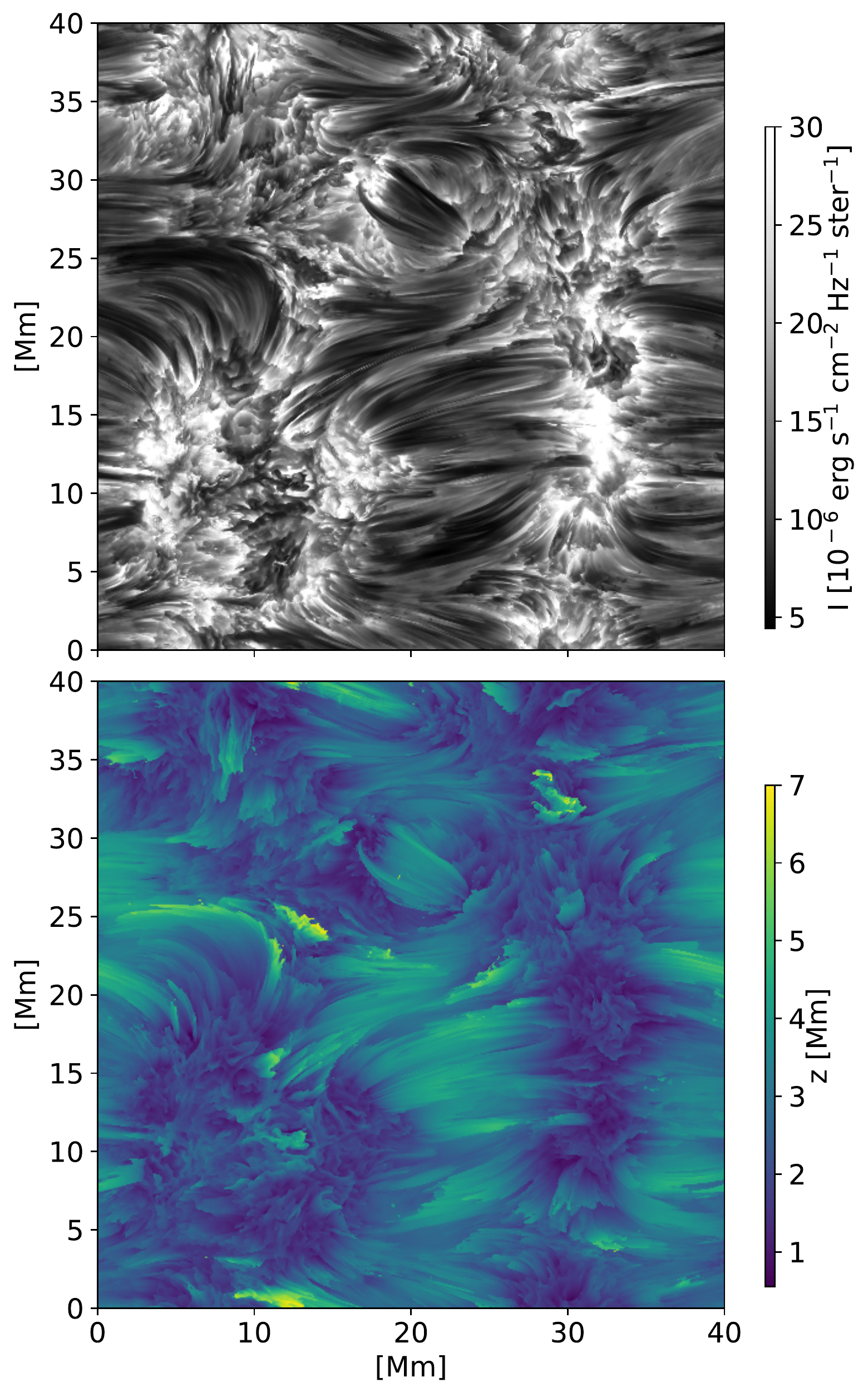}
  \caption{Vertically emergent H$\alpha$ intensity at blue wing corresponding to v$_{\rm dopp}=-36$~km\,s$^{-1}$ (top) and the line core (middle) and formation height of H$\alpha$ line core (bottom). A movie of the H$\alpha$ line-core emergent intensity is available as supplementary material.\href{https://dubshen.astro.su.se/~sdani/espm/plage_lcp.mp4}{[Movie]} }
  \label{fig:iezt1}
\end{figure}
\begin{figure}
\centering
\includegraphics[width=0.8\linewidth,trim= 0.3cm 0cm 0cm 0.2cm,clip=true]{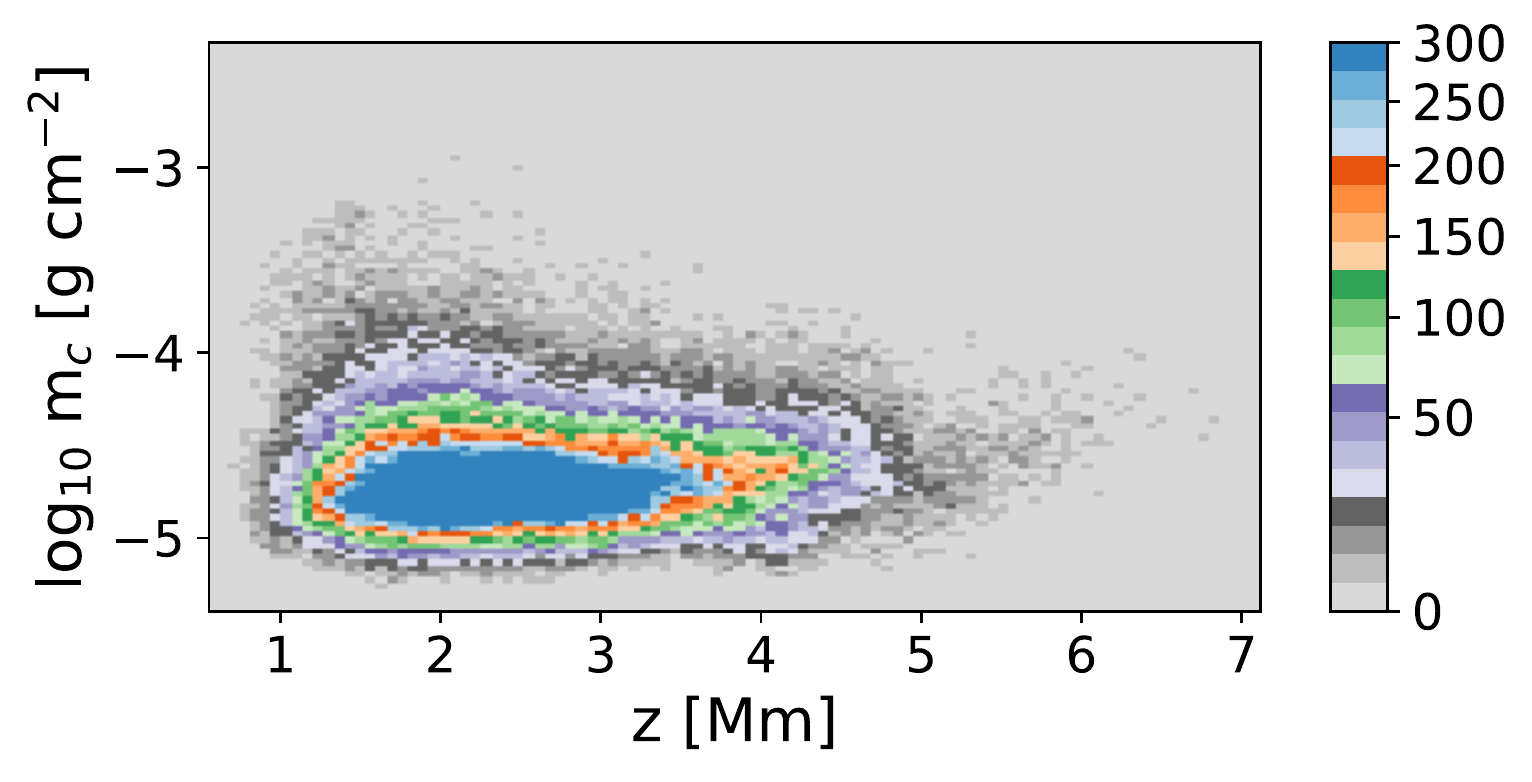}
\caption{PDF of the column mass at the H$\alpha$ line-core $\tau = 1$ height as a function of the $\tau = 1$ height. }
\label{fig:cm_vs_z}
\end{figure}

Apart from long fibrils, the chromosphere shows dynamic structures originating at fibril footpoints in the magnetic network
 \citep{2012SSRv..169..181T}. The state-of-the-art observations show that we can find flows, waves, and shocks in the chromosphere \citep{2015SSRv..190..103J,2021ApJ...920..125M,2022arXiv220301688M}. These phenomena contribute to the heating of the upper layers. Furthermore, the magnetic field and its interaction with convective flows determine how much energy is available and how it is transported further up. As a consequence, we measure different responses of different layers in the solar atmosphere depending on the observed region. This is mirrored in a varying relationship between the magnetic field and emission in diagnostics at different spectral ranges, from EUV to millimeter range \citep{2009A&A...497..273L, 2018A&A...619A...5B}. Particularly, the chromospheric temperature seems to correlate the most with the horizontal component of the chromospheric magnetic field \citep{2018A&A...612A..28L}.

This publication does not attempt to provide a review but to reflect the author's opinion on the modeling of the solar chromosphere and serves as a follow-up to a talk given at the 16th European Solar Physics Meeting. The main focus is on data-inspired models that self-consistently consider the near-surface layers of the convection zone where the actual driving of most chromospheric phenomena occurs. These models aim to reproduce the observed evolution of the magnetic field by replicating the overall magnetic field topology and flux emergence. As such, they allow studying the energy transport between subsurface layers, photosphere, chromosphere, and corona. 
\begin{figure}
  \centering
  \includegraphics[width=0.88\linewidth,trim= 0.0cm 11cm 0.2cm 0.2cm,clip=true]{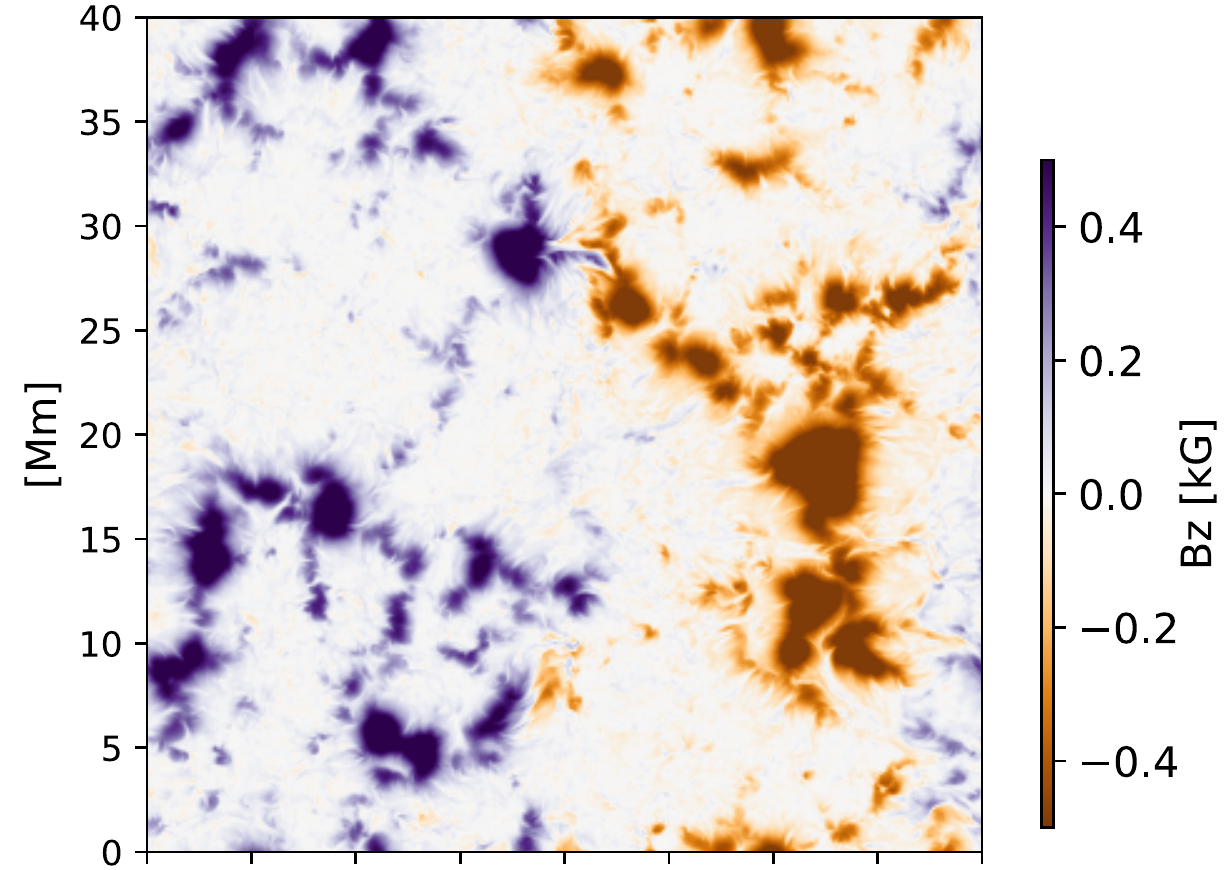}
    \includegraphics[width=0.88\linewidth,trim= 0.0cm 0cm 0.2cm 10.2cm,clip=true]{figs/bz_321000p.pdf}
  \caption{Vertical magnetic field at temperature minumum (top) and the field inclination at z$=2.7$~Mm (bottom). }
  \label{fig:bb}
\end{figure}

\begin{figure}
\centering
\includegraphics[width=\linewidth,trim= 0cm 0cm 0cm 0cm,clip=true]{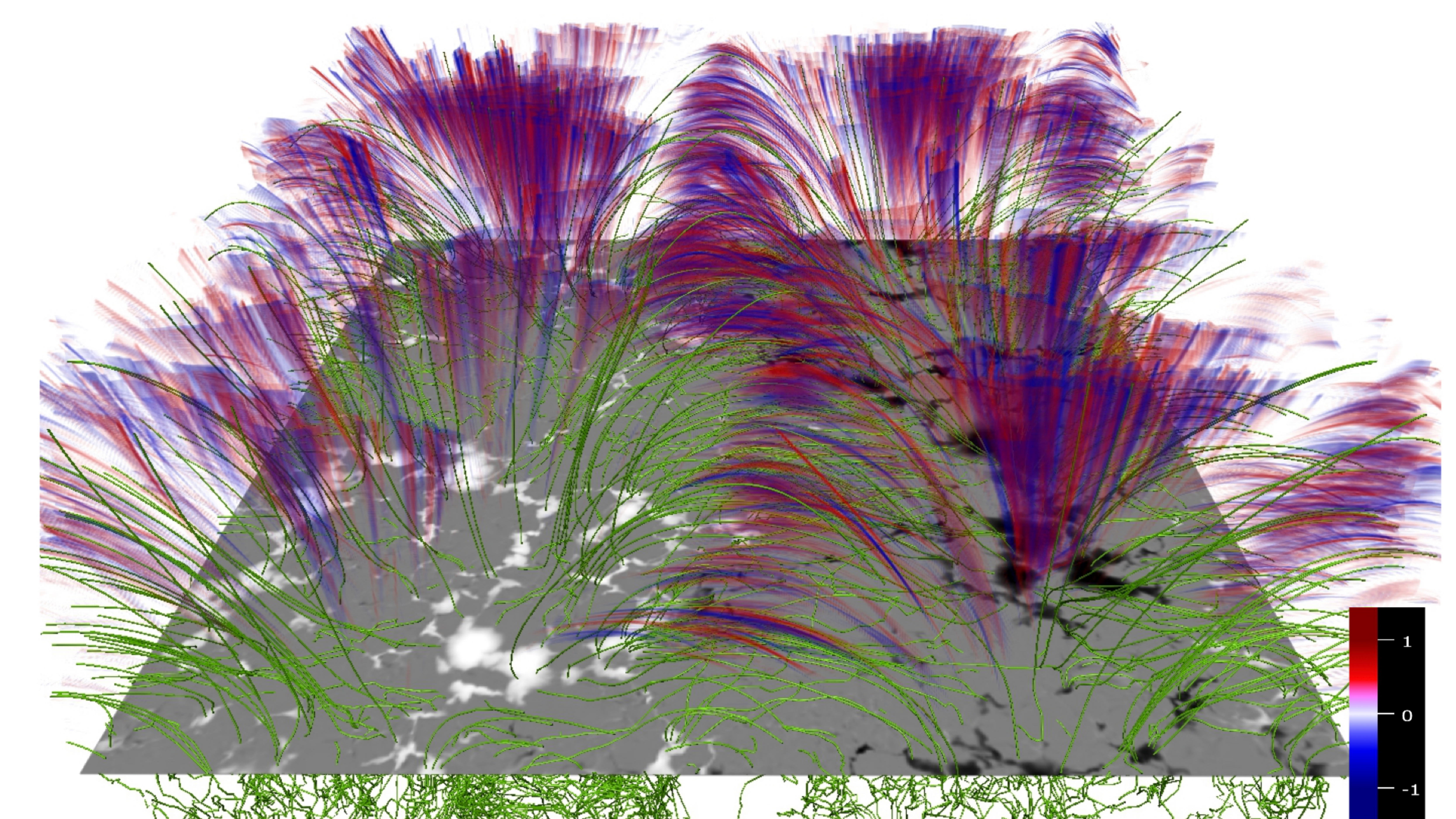}
\caption{Volume rendering of the Alfv{\'e}n wave proxi $f_{\rm alf}$ (in red and blue). The horizontal plane shows the vertical component of the magnetic field in the photosphere. The green lines outline the direction of the magnetic field lines. The figure is produced with VAPOR \citep{2019Atmos..10..488L}. }
\label{fig:Alfven_vapor}
\end{figure}

\section{Magnetic field configuration}

Features visible at the photospheric layers, at any moment in time, from small magnetic concentrations to sunspots, show different amounts of magnetic flux per unit area. Their small-scale appearance and short-temporal dynamics are a result of different regimes of so-called magneto-convection \citep[e.g.][]{2013IAUS..294...95S}. Consequently, the energy input to the higher layers differs from one region to the next.

Figure~\ref{fig:iezt1} shows a snapshot from a numerical experiment that aims to reproduce the magnetic properties of the observed plage quantitatively. The model is made by using the MURaM code \citep{2005A&A...429..335V, 2017ApJ...834...10R}. It contains only the minimum physics required to be called ‘comprehensive’. That is: single-fluid MHD, 3D grey LTE radiative
transfer, a tabulated LTE equation of state, Spitzer heat conduction, and optically thin radiative losses in the corona based on
CHIANTI \citep{2012ApJ...744...99L}. The model is built in phases starting from the hydrodynamic convection simulation that includes photosphere and near surface convection zone down to $-8$~Mm below the average $ \tau_{ \rm 500} = 1$ height. This number is a compromise. On the one hand, the domain should be deep enough to allow magnetic flux to converge toward the downflow lanes of large-scale convective cells  \citep{2009LRSP....6....2N, 2012ApJ...753L..13S}. On the other hand, the horizontal extension of the computational box should be $4$ times its depth to provide enough volume for the development of a large-scale flow pattern \citep{2013IAUS..294...95S}, but also to allow a small enough grid spacing so that small scales are reasonably resolved. In the second phase, a bipolar uniform vertical field of $200$~G is added in two halves to the fully-developed nonmagnetic convection simulation.  The initial mean unsigned flux density of $200$~G was enough to eventually form pore aggregation and magnetic structures that extended $>5$~Mm in radius. In the third phase, the potential field extrapolation is used to extend the computational domain to $14$~Mm above the average $ \tau_{\rm 500} = 1$ height and include the upper solar atmosphere. At the bottom boundary, a horizontal field at roughly equipartition field strengths is allowed to 
emerge into the domain \citep{2014ApJ...789..132R}. At the upper boundary, the field is set to be vertical. Also, the upper boundary
is open to outflows but closed to inflows. The horizontal boundaries are periodic. In the last phases, the model is first
run at a lower resolution and then again at the double resolution until a relaxed state is achieved. The final grid spacing is $39$ and $21$~km in the horizontal and vertical directions, respectively.

The middle panel in Fig.~\ref{fig:iezt1} shows emergent intensity in the H$\alpha$ line core. The synthetic images are generated with the 3D radiative transfer code Multi3D \citep{2009ASPC..415...87L}. The same H~I model atom as in \cite{2019A&A...631A..33B} is used with the LTE electron densities. The H$\alpha$ synthetic image in Fig.~\ref{fig:iezt1} shows long fibrils that spread everywhere, connecting the large features of opposite polarity outlined in the top panel of Fig.~\ref{fig:bb}. The middle and bottom panels in Fig.~\ref{fig:iezt1} can be compared with Fig.7 in \citet{2012ApJ...749..136L}. The difference between the model used here and the model used in \citet{2012ApJ...749..136L} is manifold. First, the domain of the simulations presented in this article is almost twice the size of the model used by \citet{2012ApJ...749..136L}. This allows fibrils of longer horizontal extension to be formed. The mean 
unsigned field here is $180$~G versus $30$~G in the other study. The resulting appearance of fibrils agrees with observations that show dependence on unsigned magnetic flux density. Finally, \citet{2012ApJ...749..136L} used a model with a simple bipolar configuration, while the present model has a more complex field allowing the opposite polarities to be connected in multiple directions and fibrils to cover almost the whole horizontal extent of the computational domain. The comparison of the two figures reveals that the H$\alpha$ spectral line forms in the same way in both models. The line-core image shows that the intensity is anti-correlated
to the average formation height. This means that the darker fibrils are, the higher they are formed.

Figure~\ref{fig:cm_vs_z} shows that the $\tau = 1$ height has a constant column mass of roughly $3 \times 10^{- 5}$. This trend is the same as reported in \citet{2012ApJ...749..136L}. Compared to Fig.~12 in the same article, the plot also shows branches toward the higher column masses and the higher heights. The number of occurrences in these branches is very low compared to the main cloud that is confined between $1-5 \times 10^{- 5}$~g\,cm$^{-2}$. The main cloud is shifted toward higher heights and extended to nearly $5$~Mm. This clearly shows that the same column masses are reached at significantly higher heights, meaning that much more material is lifted higher than in the QS case presented in \citet{2012ApJ...749..136L}. 

The synthetic H$\alpha$ image in Fig.~\ref{fig:iezt1} shows more than just long  fibrils. This particular snapshot shows two surges at $[12,25]$~Mm and $[30,33]$~Mm and a jet at $[11,1]$~Mm. These features are visible as protrusions in the bottom panel of Fig.~\ref{fig:iezt1} where the formation height reaches $7$~Mm or so. The preliminary analysis suggests that surges appear during magnetic field reconfiguration without a bulk flux emergence or through a mechanism similar to the one described in \cite{2017ApJ...848...38I}. 
The model also shows features that resemble short and long dynamic fibrils. Short dynamic fibrils can be visible in the region with less flux, in the lower-left quadrant of the domain where the field is mainly vertical, as the bottom panel of Fig.~\ref{fig:bb} shows. Long dynamic fibrils are visible around $[7,25]$~Mm, where the field is inclined and forms a slope. This agrees well with observations of the dynamic fibrils \citep{2007ApJ...655..624D}. Finally, the same model shows the ubiquitous presence of so-called rapid blue-shifted excursions \citep[RBE,][]{2009ApJ...705..272R} visible in the top panel of Fig.~\ref{fig:iezt1}. 

\section{Small-scales dynamics}
The second important ingredient in chromospheric modeling is the small-scale dynamics. On the smallest scales, the observations show signatures of turbulent dynamo action \citep{2010A&A...513A...1D} that results in 
mixed-polarities. Such structures can undergo a magnetic reconnection that can trigger mass loading into the chromospheric fibrils. Furthermore, the turbulent motion of the plasma sets the properties of the photospheric driver that excites magnetic footpoints. This generates different types of waves that can transport the energy to the chromosphere and higher. For example, vortex flows can serve as drivers of torsional Alfv{\'e}n waves \citep{2013ApJ...776L...4S,2021A&A...649A.121B}. Depending on the amount of magnetic flux, the observations show horizontal \citep{2010ApJ...723L.180S} and vertical kind \citep[e.g.][]{2010ApJ...723L.139B} of vortex flows present everywhere in the photosphere and chromosphere \citep{2012Natur.486..505W, 2022A&A...663A..94D}. Their presence can be recognized in specific heating signatures generated in the upper photosphere \citep{2012A&A...541A..68M}. Their abundance indicates that their contribution to coronal heating might be significant.

\begin{figure*}
\centering
\includegraphics[width=\linewidth,trim= 0.2cm 0.2cm 0cm 0.2cm,clip=true]{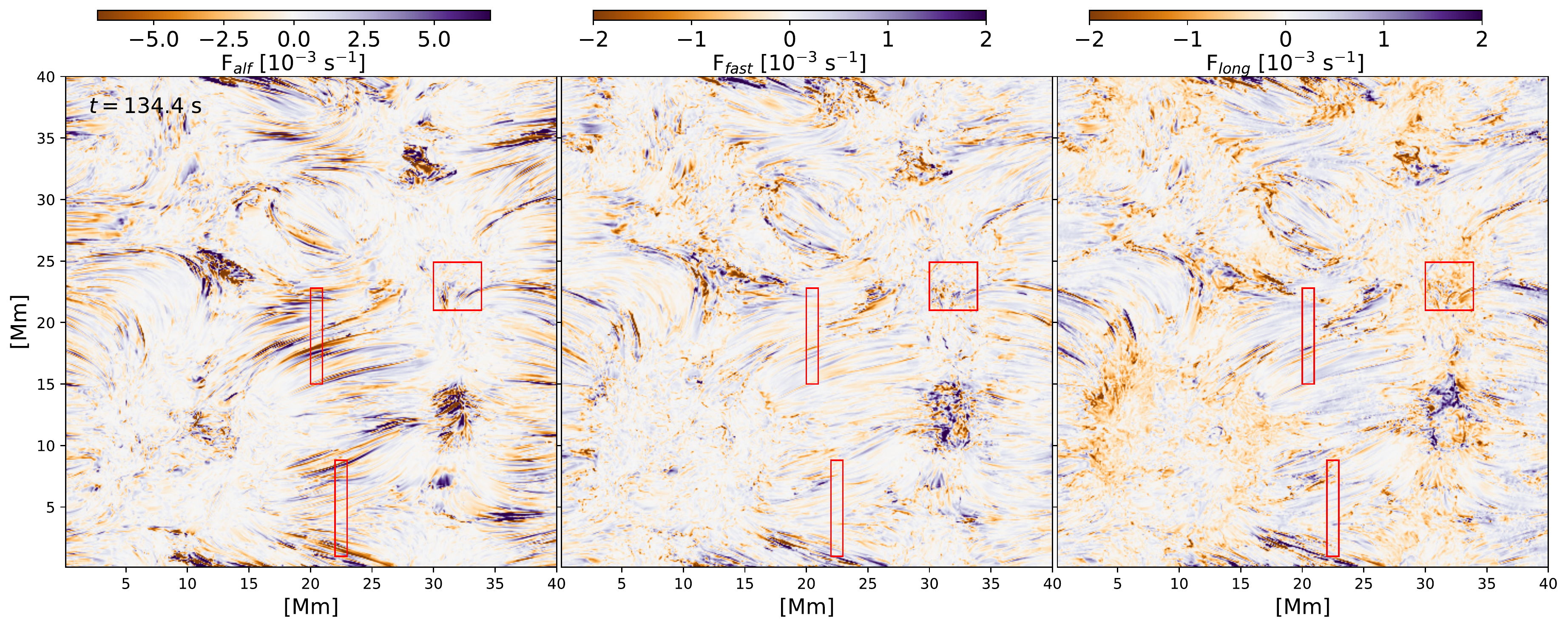}
\caption{Proxies for Alfv{\'e}n $f_{\rm alf}$ (left), fast $f_{\rm fast}$ (middle) and slow magneto-acoustic $f_{\rm long}$ (right) waves averaged over the the formation height of the H$\alpha$ line core. \href{https://dubshen.astro.su.se/~sdani/espm/fwave.mp4}{[Movie]} The red rectangles mark the regions that are used for power spectra averaging shown in Fig.~\ref{fig:power}. }
\label{fig:fwave}
\end{figure*}

The question remains to what extent the present simulation can resolve and reproduce plasma dynamics on the smallest scales. When it comes to waves, in the complex simulations like the ones presented in the previous section, identifying and following different wave modes is difficult. To separate them and analyze their spatial and temporal characteristics, the following quantities are calculated after \cite{2018A&A...618A..87K}:

$f_{\rm alf}= \hat{e_{\parallel}}\cdot \bigtriangledown \times \mathbf{v}$ 

$f_{\rm fast}= \bigtriangledown \cdot \left ( \mathbf{v} -  \hat{e_{\parallel}}v_{\parallel} \right )$ 

$f_{\rm long}= \hat{e_{\parallel}} \cdot \bigtriangledown \left ( \mathbf{v} \cdot \hat{e_{\parallel}} \right )$  

where $\mathbf{v}$ is the velocity vector and $\hat{e_{\parallel}}$ the field-aligned unit vector. The quantities represent proxies for three wave modes slow $f_{\rm long}$ and fast $f_{\rm fast}$ 
magneto-acoustic and Alfv{\'e}n $f_{\rm alf}$. Figure~\ref{fig:Alfven_vapor} shows distribution of the quantity $f_{\rm alf}$ for the snapshot shown in Fig.~\ref{fig:iezt1}. Logically, the quantity gets larger with height and follows the magnetic field orientation. To visualize where the wave power for each mode is located, each quantity is averaged within the layer where column mass ranges $-5< \log_{10}(m_{\rm c}) < -4.5$. As Fig.~\ref{fig:iezt1} shows, this is the chromosphere as we see it in the H$\alpha$ line core. The resulting maps are shown in Fig.~\ref{fig:fwave}. The quantity $f_{\rm alf}$ is predominant in regions with the horizontal field. High $f_{\rm alf}$ quantities can also be found in surges and the strong magnetic feature at $[32,13]$~Mm. The other two quantities are not as strong. They have a morphology similar to $f_{\rm alf}$ along the closed loops. In regions with open or inclined field, the distribution of $f_{\rm alf}$  and  $f_{\rm fast}$  is similar. The distribution of $f_{\rm long}$, unlike the other two quantities, shows increased value along the rims between open and closed field regions, i.e. at the loop footpoints. The supplementary movie shows the temporal evolution of these quantities over $635$~s with the cadence of $\sim 10$~s. Rapid changes are visible in all three quantities. In most places, the quantities show abrupt changes in sign and magnitude. The changes seem faster in regions with the closed field. In some places, an apparent propagation can be identified. Proxies of all three wave modes are detectable whenever a more explosive event happens i.e. onset of a small flare at the end of the time series at $[27,22]$~Mm and $[38,36]$~Mm.

\begin{figure}[h!]
\centering
\includegraphics[width=\linewidth,trim= 0.2cm 2.5cm 0cm 0.2cm,clip=true]{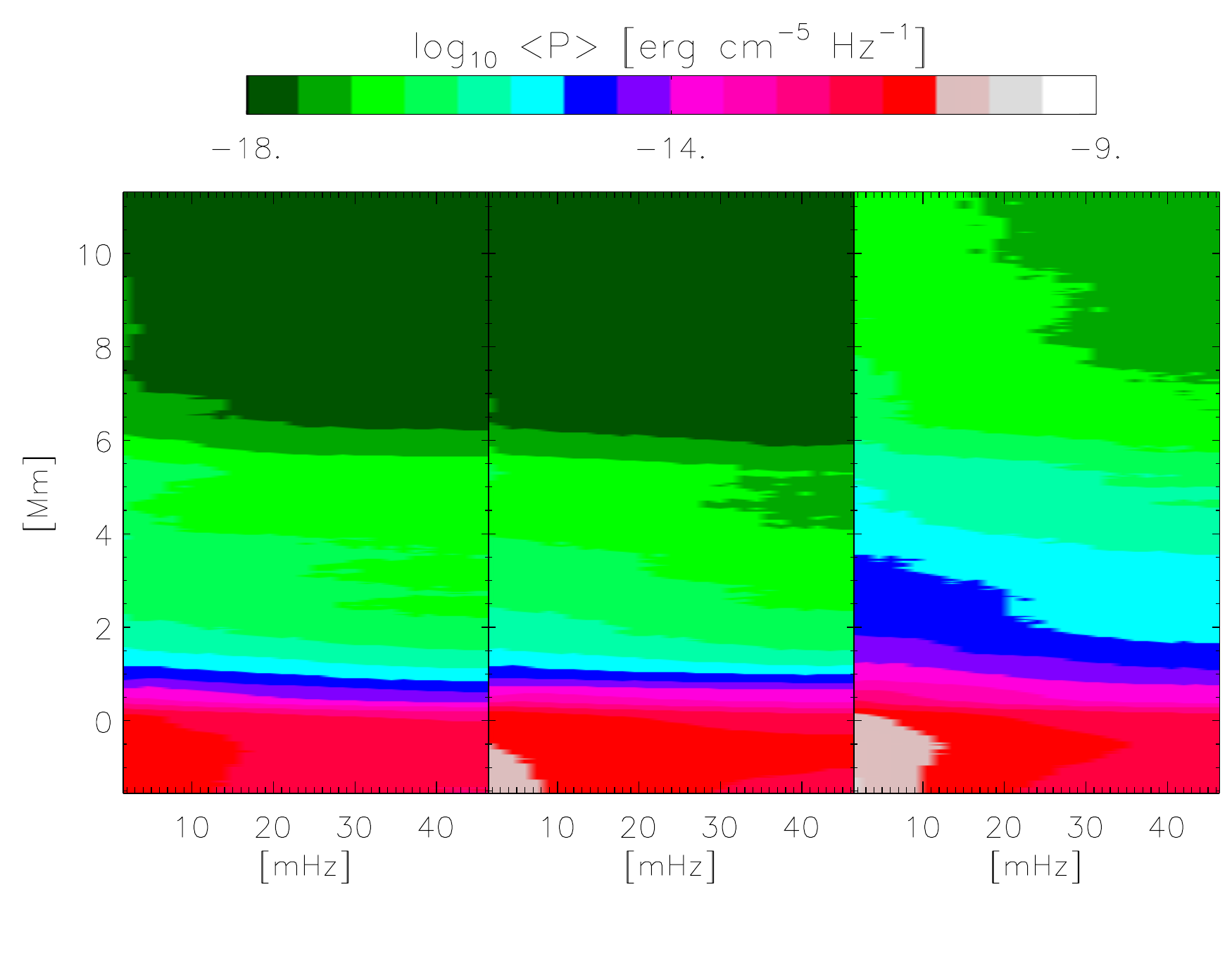}
\includegraphics[width=\linewidth,trim= 0.2cm 1.2cm 0cm 2.5cm,clip=true]{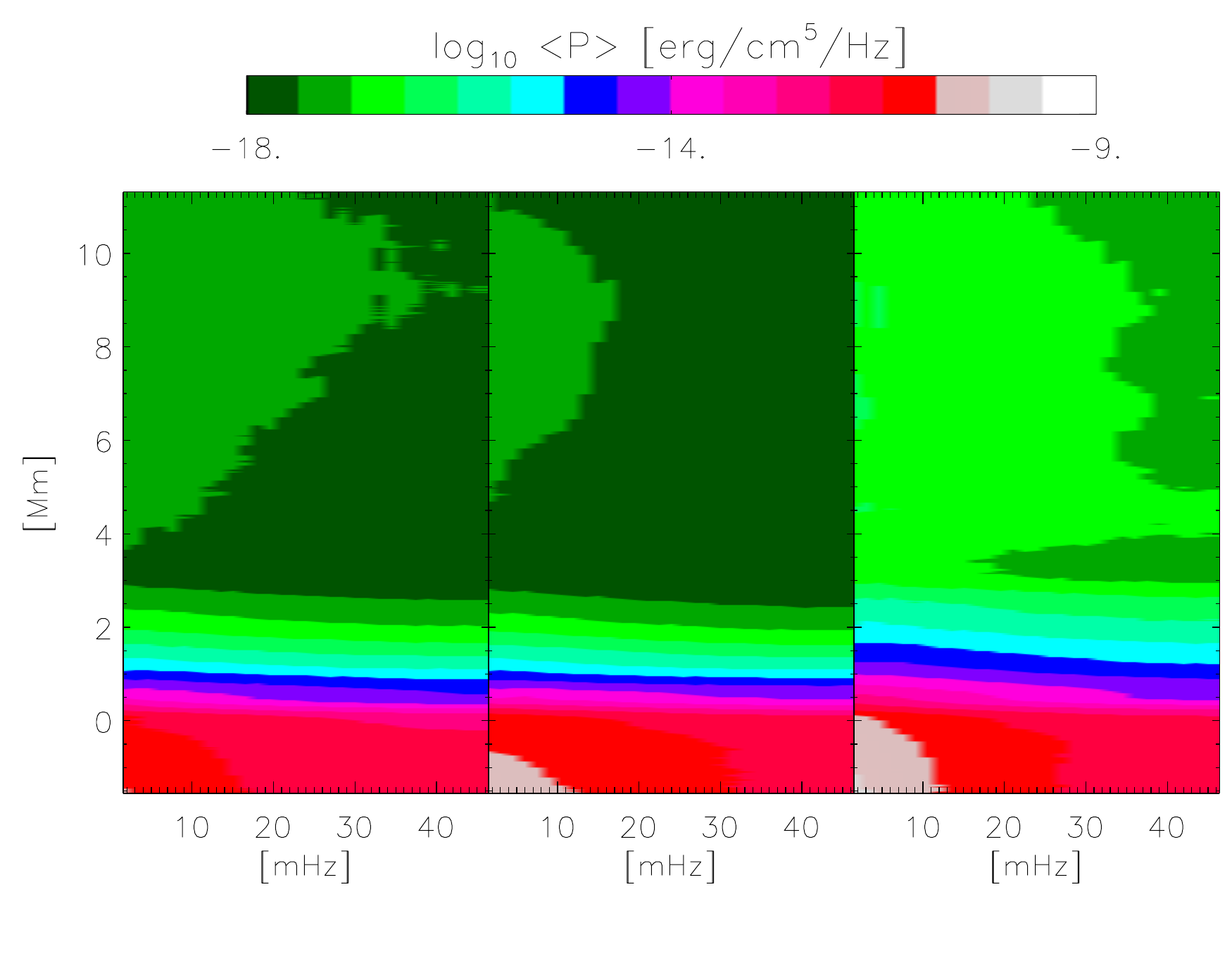}
\caption{Power spectra of $f_{\rm long}$  (left), $f_{\rm fast}$ (middle) and $f_{\rm alf}$ (right) as a function of height and frequency in regions with horizontal (top) and vertical (bottom) field.}
\label{fig:power}
\end{figure}

The three outlined regions are used to produce averaged power spectra shown in Fig.~\ref{fig:power}. The red quadrant outlines a region with open/inclined field. The two rectangles cover high canopy and, when summed, outline region of equal size. Averaged power spectra are multiplied with the corresponding average density profile and plotted as a function of height and frequency.  

The power maps for two field configurations are significantly different. The differences are evident in all three quantities. The power over all frequencies decreases less rapidly for horizontal field configuration (top row of Fig.~\ref{fig:power}) and stays constant in the range of heights between $4-6$~Mm where chromospheric canopy coincides (see bottom panel of Fig.~\ref{fig:iezt1}). Only $f_{\rm alf}$ shows higher power extending further up to $11$~Mm height and only at frequencies lower than $20$~mHz. In the case of a more vertical field configuration (bottom row of Fig.~\ref{fig:power}) the power drop off is gradual with height. All three quantities show a higher power in the frequency range between $2$ and $10$~mHz all the way up to the corona. The leakage towards higher frequencies is present for both $f_{\rm long}$ and $f_{\rm alf}$, but over different height ranges.

The power distribution shown in Fig.~\ref{fig:power} is in agreement with several observational studies. There are many reports on the detection of oscillations with periods around $5$~minutes in solar corona \citep[][and references therein]{2005ApJ...624L..61D}. They are mostly interpreted as upward-propagating longitudinal magnetoacoustic
waves, but there are reports that state detection of Alfv{\'e}n \citep{2007Sci...317.1192T} or transverse kink waves \citep{2008A&A...487L..17V} which should be fast magnetoacoustic waves \citep{2012RSPTA.370.3193D}. Detecting oscillations at high frequencies is, however, harder primarily due to the limited number of photons available in the UV/EUV part of the solar spectrum, as well as in strong chromospheric lines. There are, however, a few studies based on the H$\alpha$ narrow-band filtergrams. \cite{2021ApJ...921...30S} analyzed the sudden appearance and disappearance of dark features in H$\alpha$ wings. They found high-frequency oscillations with an average frequency between $10-63$~mHz which they interpreted as fast magnetoacoustic waves. Similarly, \cite{2011ApJ...736L..24O} and more recently \cite{2022ApJ...930..129B} studied transverse oscillations in spicules, features visible off the disk. They detected high-frequency waves with periods down to $10$~s named Alfv{\'e}nic, which allows them to be both Alfv{\'e}n or fast magnetoacoustic waves in nature.  



\section{Effect of the emerging flux}
The third important ingredient for reproducing chromospheric features and dynamics is taking into account the flux emergence. Emerging flux has a significant role in filling the Sun’s atmosphere with hot plasma. The flux emergence happens on a wide range of spatial scales \citep{2011SoPh..269...13T,2016ApJ...820...35G,2017ApJS..229...17S}. Although most of the flux emerging on granular scales gets pulled back \citep{2010ApJ...723L.149D}, some percentage rise and may contribute to the heating of higher atmospheric layers \citep{2009ApJ...700.1391M}. Reconnection of the newly emerging with the pre-existing magnetic field may be an important source of (1) the heating and filling of the coronal loops with mass \citep{2017ApJS..229....4C} (2) generation of jets on different spatial scales \citep{1996PASJ...48..353Y} (3) generation of spicules \citep{2019Sci...366..890S} and so on.  Localized transient heating events are most likely always related to emerging magnetic flux which brings substantial magnetic and kinetic energy at short time scales. It is suggested by observations \citep{2014Sci...346C.315P} and also confirmed with numerical models \citep{2022A&A...661A..59D} that transition region and coronal temperatures can be reached at much lower heights in the atmosphere when violent events take place. Finally, on the larger scales, observations show high correlations of the temperature with the strength of the horizontal magnetic field in the low chromosphere of the emerging regions \citep{2018A&A...612A..28L}. 

The following numerical experiments illustrate how the different properties of emerging flux result in different heating rates in the chromosphere and the formation of different chromospheric features. The two experiments are a continuation of the run presented in section $2$. They start with the same field configuration but differ in properties of the bipolar flux system that is advected through the bottom boundary within an area with an elliptical shape \citep{2019NatAs...3..160C}. In the first case, flux emergence run 1 (FER1), the emergence is limited within an ellipsoidal area with a major and minor axis 
$(a, b) = (10, 2)$~Mm and field strength of $5000$\,G. In the second, FER2, the emerging flux is confined within $(a, b) = (3, 1)$~Mm and field strength of $8000$\,G. So FER1 has less flux advected over a larger area and FER2 has the opposite. As a result, the emerging flux in two numerical experiments reaches the photosphere at different times. Due to the difference in properties of the emerging flux, the two experiments lead to different magnetic field evolution as visible in the top row of Fig.~\ref{fig:emergence_snap} and in supplementary movies. Time $t=0$ is arbitrarily chosen for both experiments and marks the moment when emerging flux reaches the photosphere.

The FER1 was inspired by the observations analyzed in \cite{2018A&A...612A..28L}. The aim was to reproduce the orientation of the emerging with respect to the preexisting field. The major feature of this run is a peacock jet that gets generated as the emerging flux pushes into the large negative polarity at $[35,23]$~Mm and forms a positive intrusion. Observations that inspired the numerical experiment also show a peacock jet whose properties are greatly reproduced in the model. The peeling of magnetic field lines due to reconnection with the emerging field as well as subsequent collimation of the ejected material differs from the scenarios presented in \cite{1996PASJ...48..353Y} and \cite{2013ApJ...771...20M}. 

The FER2 is partly analyzed in \cite{2022A&A...661A..59D}. That analysis focuses on the snapshot at $t=36.6$~min that produced the greatest brightness excess in the synthetic $3$~mm emission. As that study shows, the 3 mm
emission largely outlines dissipation in current sheets that are generated in the interaction region between emerging and pre-existing systems. Some $10$~min after the analyzed moment, there is a flare onset.

Bottom panels of Fig.~\ref{fig:emergence_snap} show the sum
of the viscous and resistive heating rates integrated over the height that corresponds to column mass in the range $-5< \log_{10}(m_{\rm c}) < -4.5$. The difference between the runs is largest just above the emergence region $[12-27,15-30]$~Mm. In the case of FER1, the heating patterns are mostly in form of stripes in the direction of overlying field lines. The heating is of the order of $20$~kWm$^{-2}$ in these cases. There are cases of localized heating that exceed $80$~kWm$^{-2}$ and these events resemble the UV bursts \citep{2014Sci...346C.315P}. In the case of FER2, localized heating events are larger and volume-filling, with heating rates always larger than $100$~kWm$^{-2}$. This is in line with values extracted from observations \citep{2021A&A...647A.188D}. With the onset of the flare, the heating of $> 100$~kWm$^{-2}$  covers the whole area above the emerging flux.

\begin{figure}
\centering
\includegraphics[width=0.47\linewidth,trim= 0.2cm 0.32cm 3cm 0.35cm,clip=true]{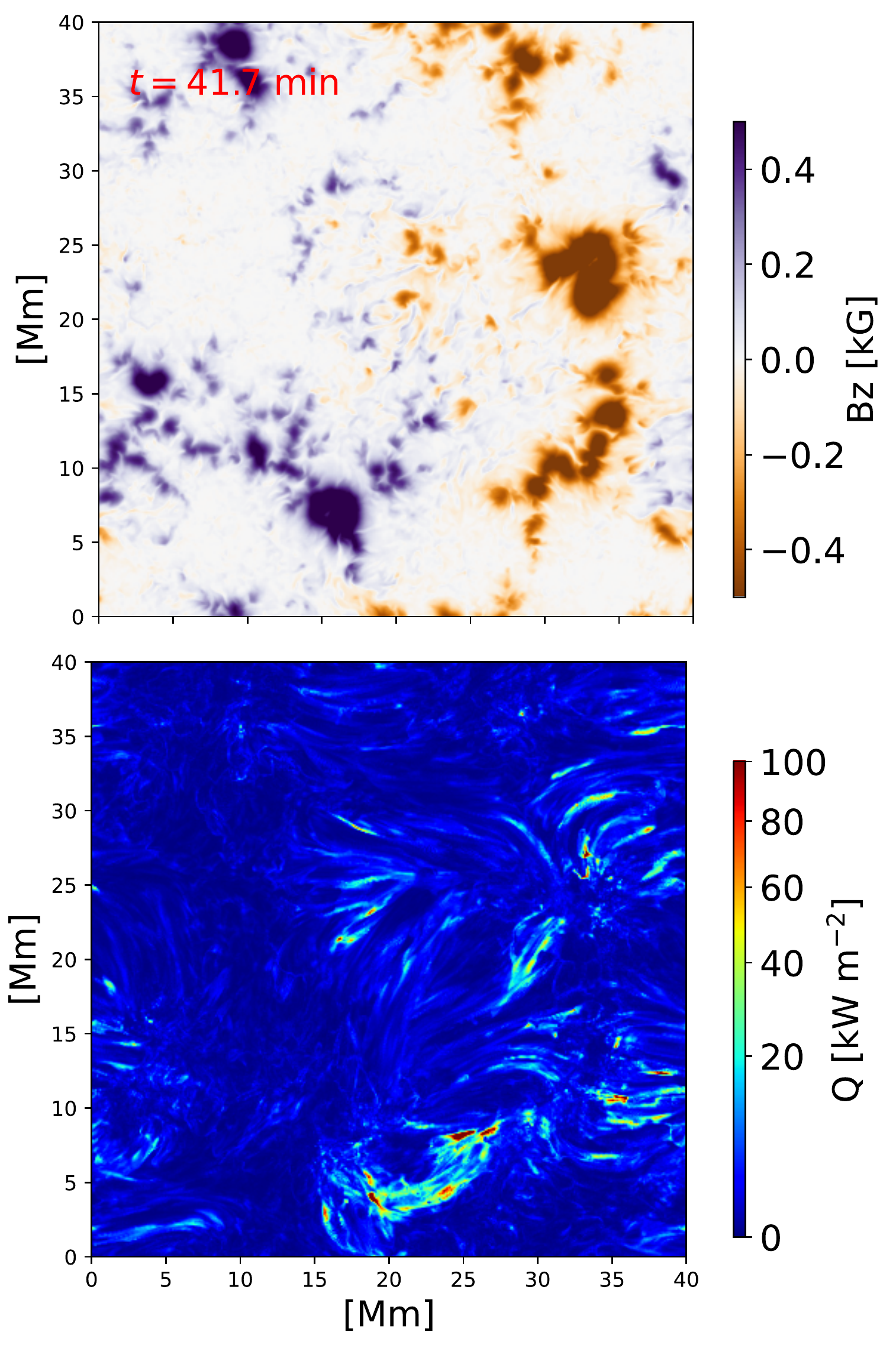}
\includegraphics[width=0.52\linewidth,trim= 1.5cm 0.32cm 0.38cm 0.35cm,clip=true]{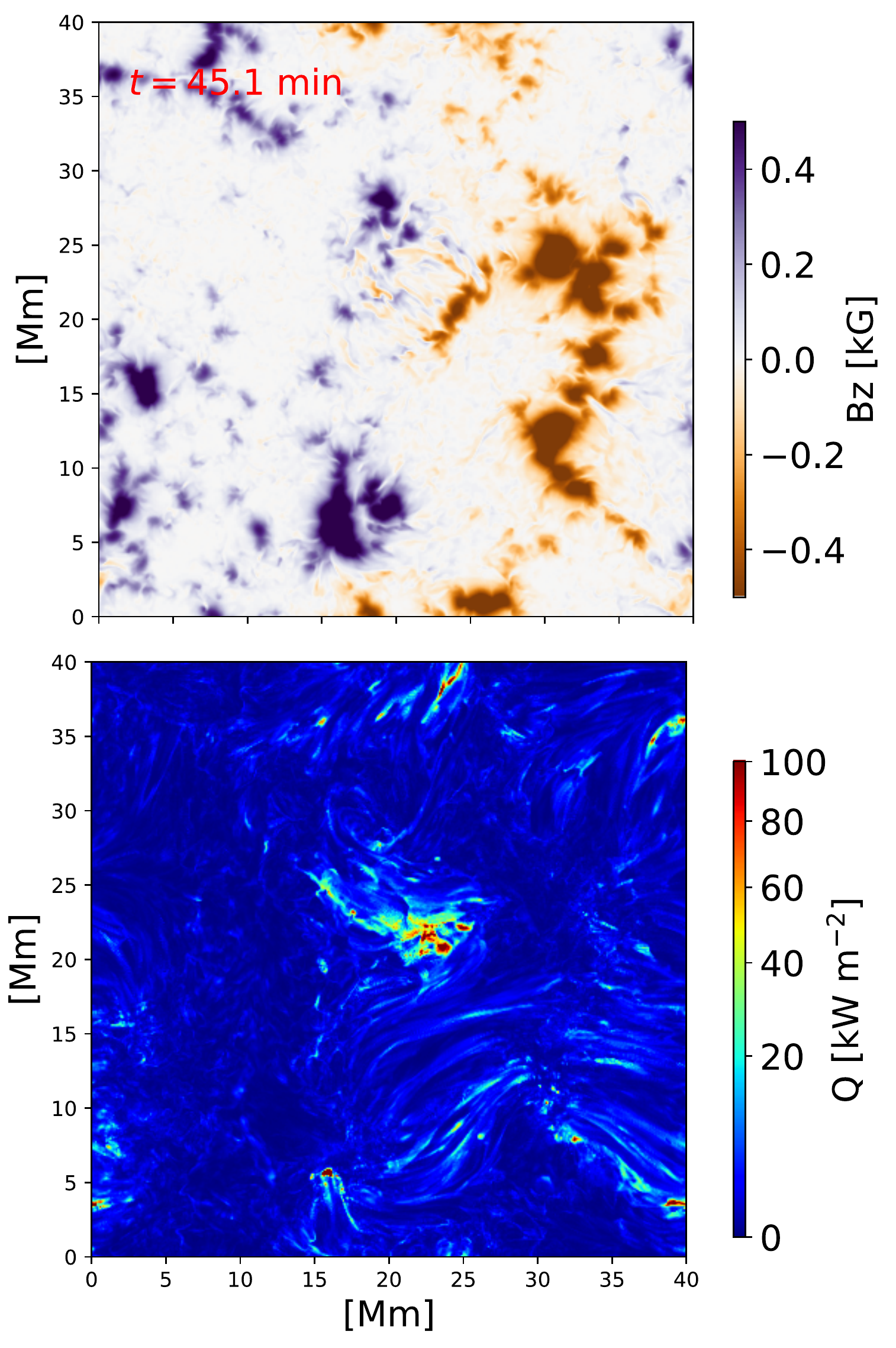}
\caption{Snapshots from two runs that differ in emergence flux advected into the computational domain (left- FER1 and right - FER2). Top row - vertical component of the magnetic field. Bottom row - Sum of resistive and viscous heating terms summed over column masses $-5< \log_{10}(m_{\rm c}) < -4.5$. Movies are available as supplementary material.\href{https://dubshen.astro.su.se/~sdani/espm/losses_FER1.mp4}{[FER1 movie]} \href{https://dubshen.astro.su.se/~sdani/espm/losses_FER2.mp4}{[FER2 movie]} }
\label{fig:emergence_snap}
\end{figure}

\section{Computationally expensive physics}
Great efforts are made in building models that go beyond the single fluid description. A multi-fluid treatment that separates ion and neutral species is necessary for some atmospheric processes \citep{2018SSRv..214...58B}. This comes from the fact that the ionization fraction increases rapidly with height which makes the chromosphere essentially collisionally weakly coupled. The easy way to extend the single-fluid model to incorporate the two-fluid plasma effects is by adding the two additional terms in generalized Ohm’s law. These are ambipolar diffusion (AD) with the Hall effect \citep[e.g.][]{2021ApJ...923...79R}. However, as shown by \cite{2020A&A...633A..66N}, the inclusion of these effects does not make sense without also including the non-equilibrium ionization of hydrogen. The two must be taken into account together because including only AD with the assumption of LTE underestimates the ionization fraction and as a result, overestimates the effect of ambipolar diffusion. 

Although the numerical experiments presented in previous sections do not account for any computationally expensive physics, they do manage to reproduce many chromospheric features and their dynamics to a large extent. This is in disagreement with previous studies that state that the generation of features observed in the chromosphere, i.e. long fibrils and spicules is not possible without the inclusion of non-equilibrium ionization of hydrogen and ambipolar diffusion \citep{2017Sci...356.1269M, 2019A&A...632A..96R, 2019ARA&A..57..189C}. The models presented here show even quantitative agreement with observed values \citep{1977ARA&A..15..363W} at the spatial resolution presently available
 \citep{2022A&A...661A..59D}. This comes as a surprise since numerical diffusivity and viscosity terms are larger than their real physical values. However, the numerical experiments show that the Poynting flux input in the photosphere determines the total energy dissipation rate while variations in numerical diffusivity and viscosity terms have only a marginal effect  \citep{1996JGR...10113445G,2017ApJ...834...10R}. 
 
Including both non-equilibrium ionization of hydrogen and the ion-neutral interaction effects are important for several reasons. The dynamics and temperature in the upper chromosphere and transition region will be largely affected. The resulting temperature variation will be finer and small-scale pockets with higher density will be formed \citep{2020A&A...633A..66N,2020ApJ...889...95M}. As a result, also the appearance of chromospheric features from long fibrils to spicules and dynamic fibrils will be altered. For example, it is demonstrated that H$\alpha$ line core intensity outlines density ridges and the line-core width correlates with temperature \citep{2009A&A...503..577C,2012ApJ...749..136L}. The aforementioned changes in density and temperature distribution would lead to larger variations in H$\alpha$ line intensity and this may result in e.g. apparently less opaque long fibrils than shown in Fig.~\ref{fig:iezt1}. Furthermore, the heating through ion-neutral collisions will affect the reconnection rate and may give more homogenous heating in the emergence regions  \citep{2006A&A...450..805L}. Finally, the inclusion of ambipolar diffusion and the Hall effect will further help the production of Poynting flux in the form of Alfv{\'e}n waves \citep{2021RSPTA.37900176K}.

\section{Summary}
As \cite{1977ARA&A..15..363W} write: 'The configuration of the magnetic field is an important factor in controlling the energy losses.'. So to be able to properly model the solar atmosphere and with it the most complex layer, chromosphere, we have to start from there. The final recipe for building ideal atmospheric models of solar plage starts with reproducing magnetic field configuration. The first phase would be reproducing photospheric magnetic features e.g. their size and the field strength. We can adjust this by choosing correctly the strength and configuration of the initial field and setting the lower boundary of the computational domain sufficiently deep so that the magnetic flux can converge and create extended magnetic features as we observe in the Sun. By choosing all these parameters right we are self-consistently reproducing the correct regime of magneto-convection and ensuring that the available energy input is correct. The second phase is choosing the correct time to extend the domain to the chromosphere. The correct overall magnetic field configuration then provides channels for energy and mass to be transferred to upper layers the same way as it might happen on the Sun. The final phase is introducing the emerging flux if we see its signatures in the observations.  

Once the overall magnetic field configuration matches the observations, we can add computationally expensive physics: non-equilibrium ionization of hydrogen and ambipolar diffusion with the Hall effect. This final step would possible close the gap between observation and models. The only ingredient that is missing then is the effect of non-thermal particles. Although efforts are made to include the effect of the non-thermal particles into realistic MHD simulations  \citep{2018MNRAS.477..624N, 2020A&A...643A..27F, 2020ApJ...896...97R}, these models are still computationally expensive to be used for one-to-one comparison with observations on the larger scales. By building one-to-one models, we can identify the exact location where the effect of non-thermal particles may be significant. This will be visible in the differences between observed and synthetic observables. In this way, one-to-one models will complement RADYN models \citep{1992ApJ...397L..59C} that do not take into account the magnetic field and its forces acting on plasma.

\section{Acknowledgments}
S.D. thanks Matthias Rempel, Johan Bj{\o}rgen and Jorrit Leenaarts for sharing their codes which allowed these results to be obtained. This project has received funding from Swedish Research Council (2021-05613), Swedish National Space Agency (2021-00116) and the Knut and Alice Wallenberg Foundation. This research data leading to the results obtained has been supported by SOLARNET project that has received funding from the European Union’s Horizon 2020 research and innovation programme under grant agreement no 824135. The calculations were performed on resources provided by the Swedish National Infrastructure for Computing (SNIC) at the National Supercomputer Centre (NSC) at Linköping University and the PDC Centre for High Performance Computing (PDC-HPC) at the Royal Institute of Technology in Stockholm. 



\end{document}